\documentclass[12pt,preprint]{aastex}

\usepackage{lscape}

\begin{document}

\shorttitle{Radial velocities from VLT-KMOS spectra in NGC~6388}
\shortauthors{Lapenna et al.}

\title{Radial velocities from VLT-KMOS spectra of giant stars in the globular cluster NGC 6388\footnote{Based on observations collected at the ESO-VLT (Cerro Paranal, Chile) under program 60.A-9448(A).}}

\author{
E. Lapenna\altaffilmark{2}, 
L. Origlia\altaffilmark{3},
A. Mucciarelli\altaffilmark{2}, 
B. Lanzoni\altaffilmark{2}, 
F. R. Ferraro\altaffilmark{2},
E. Dalessandro\altaffilmark{2},
E. Valenti\altaffilmark{4},
M. Cirasuolo\altaffilmark{5},
}
\affil{\altaffilmark{2} Dipartimento di Fisica e Astronomia, Universit\`a degli
  Studi di Bologna, Viale Berti Pichat 6/2, I--40127 Bologna, Italy}
\affil{\altaffilmark{3} INAF- Osservatorio Astronomico di Bologna, Via
  Ranzani, 1, 40127 Bologna, Italy}
\affil{\altaffilmark{4} European Southern Observatory,
  Karl-Schwarzschild-Strasse 2, 85748 Garching bei M\"unchen, Germany}
\affil{\altaffilmark{5} Institute for Astronomy, University of Edinburgh \& STFC,
 UK Astronomy Technology Center Royal Observatory, Blackford Hill, EH9 3HJ, Edinburgh, U.K.}

\date{October, 2014}

\begin{abstract} 
We present new radial velocity measurements for 82 stars,
members of the Galactic globular cluster NGC~6388, obtained from
ESO-VLT KMOS spectra acquired during the instrument Science
Verification. The accuracy of the wavelength calibration is
discussed and a number of tests of the KMOS response are
presented. The cluster systemic velocity obtained
($81.3\pm 1.5$ km s$^{-1}$) is in very good agreement with previous
determinations. While a hint of ordered rotation is found between
$9\arcsec$ and $20\arcsec$ from the cluster centre, where the
distribution of radial velocities is clearly bimodal, more data are
needed before drawing any firm conclusions. The acquired sample of
radial velocities has been also used to determine the cluster velocity
dispersion profile between $\sim 9\arcsec$ and $70\arcsec$,
supplementing previous measurements at $r<2\arcsec$ and
$r>60\arcsec$ obtained with ESO-SINFONI and ESO-FLAMES spectroscopy,
respectively. The new portion of the velocity dispersion profile
nicely matches the previous ones,  better defining the knee of the
distribution.
The present work clearly shows the effectiveness of a deployable Integral
Field Unit in measuring the radial velocities of individual stars for determining the
velocity dispersion profile of Galactic globular clusters.
It represents the pilot project for an ongoing
large program with KMOS and FLAMES at the ESO-VLT, aimed at
determining the next generation of velocity dispersion and rotation
profiles for a representative sample of globular clusters.
\end{abstract}

\keywords{globular clusters: individual (NGC 6388) -- instrumentation:
  spectrographs -- stars: kinematics and dynamics -- techniques:
  spectroscopic}

\section{Introduction}
Galactic globular clusters (GCs) are massive ($10^4-10^6 M_\odot$)
stellar aggregates, where the two-body relaxation
time-scale is shorter than the age \citep[e.g.,][]{BT87}.  For this
reason, they have been traditionally assumed to be quasi-relaxed,
non rotating systems, characterized by spherical symmetry and
orbital isotropy.  Hence spherical, isotropic and non-rotating
models, with a truncated distribution function close to a Maxwellian
\citep{king66,wilson75} are commonly used to fit the observed
surface brightness or density profiles, and to estimate the main GC
structural parameters, like the core and half-mass radii, the
concentration parameter and even the total mass
\citep[e.g.][]{harris96, mclvdm05}.

However, recent theoretical results indicate that these systems may
have not attained complete energy equipartition \citep{trenti13} and,
depending on the degree of dynamical evolution suffered and the effect
of an external tidal field, they may still preserve some
characteristic kinematical feature \citep{vesperini14}.  In
particular, non zero angular momentum has been recognized to affect
the entire dynamical evolution of star clusters \citep{einsel99}, and
central rotation might still be present in GCs hosting an intermediate
mass ($10^2-10^4 M_\odot$) black hole \citep[IMBH;][]{fiestas10}.
Moreover, it is well known that the density profile alone is not
sufficient to fully characterize a gravitational system, and the
information about internal dynamics is also necessary
\citep[e.g.][and references therein]{BT87,meyheggie97}.

For instance, a star density profile with a shallow cusp deviating from a
King (1966) model and a  velocity dispersion (VD) profile 
with a keplerian central behavior are predicted in the presence of an IMBH \citep[e.g.][]{baumg05, miocchi07}.
Despite its importance, the kinematical properties of Galactic GCs are
still poorly explored from the observational point of view, although
the number of dedicated studies aimed at building their VD and
rotation profiles has significantly increased in the last years
\citep[see, e.g., ][and references therein]{anderson10, noyola10,
lane10, bel12, mcnamara12, lut13, fabricius14, kacharov14}.
In this context, interesting insights on specific dynamical processes
occurring in the central regions of some clusters have been obtained by
using ``exotic'' stellar populations, like millisecond pulsars and
blue straggler stars \citep[see][]{fe03,fe09_m30,fe12}. However
proper VD and rotation profiles, especially in their innermost regions
where the presence of the long-searched IMBHs is expected to leave
characteristic signatures (as a VD cusp and systemic rotation;
\citealp{baumg05, miocchi07,einsel99}), are still badly
constrained. This is due to the observationally difficulties,
affecting both proper motion studies and the investigations of the
velocity line-of-sight component.

As for the latter, while determining the line-of-sight rotation curve
and VD profile in external galaxies is relatively simple (requiring
the measurement, respectively, of the Doppler shift and the broadening
of spectral lines in integrated-light spectra), it is much less
straightforward in resolved stellar populations as Galactic GCs.  In
these systems, in fact, the dominant contribution of a few bright
stars may artificially broaden the spectral lines, making the
resulting value a non-representative measure of the true VD of the
underlying stellar population \citep[this is commonly refereed as
``shot noise bias''; e.g.][]{dubath97,noyola10,lut11}. The
alternative approach is to measure the dispersion about the mean of
the radial velocities of statistically significant samples of
individual stars.  Clearly, this methodology is not prone to the
shot-noise bias, provided that the individual stars are well resolved,
sufficiently isolated and bright enough to be negligibly contaminated
by the unresolved stellar background.

The latter approach is becoming increasingly feasible thanks to the
current generation of adaptive-optics (AO) assisted spectrograph
with an integral field unit (IFU), and the improved data analysis techniques
(e.g. \citealp{lan13}, hereafter L13; \citealp{kamann13}),
as clearly demonstrated by the case of NGC~6388.
The VD profile of this cluster derived from the line broadening of
integrated-light spectra shows a steep cusp with a central value of
23-25 km s$^{-1}$, which is best-fitted by assuming that an IMBH of
$2\times 10^4 M_\odot$ is hidden in the system \citep{lut11}.
Instead, if the radial velocities of individual stars are used, a
completely different result is found.  By using SINFONI, an
AO-assisted IFU spectrograph at the ESO-VLT, L13 measured the radial
velocity of 52 individual stars in the innermost $2\arcsec$ of the
cluster, finding a flat VD profile with a central value of only 13-14
km s$^{-1}$, which is well reproduced by no IMBH or, at most, a BH of
$\sim 2000 M_\odot$ (L13; see also \citealp{lan07}).  As discussed in
detail in L13 (see their Sect. 4.1 and their Fig. 12), the integrated
light spectra measured in the innermost part of the cluster are
dominated by the contribution of two bright stars having opposite
radial velocities with respect to the systemic one, despite the
explicit effort by \citet{lut11} to correct for this. This produces a
spuriously large line broadening and a consequent overestimate of the
central VD value.

The results obtained in NGC 6388 clearly demonstrate the feasibility
of the individual radial velocity diagnostics and show that this is
indeed the safest way to measure the stellar VD in Galactic GCs. To
identify other multi-object facilities suitable for this kind of
approach, we took advantage of the new $K$-band Multi Object
Spectrograph \citep[KMOS;][]{sharples10}, recently commissioned at the
ESO-VLT.  During the instrument Science Verification (SV) run, under
proposal 60.A-9448(A) (PI: Lanzoni), we used KMOS multiple pointings
to investigate the region within $\sim 9\arcsec$ and $70\arcsec$ from
the center of NGC 6388. The results obtained from these observations
are the subject of the present paper, and they prompted us to
successfully apply for an ESO Large Program (193.D-0232, PI: Ferraro)
aimed at constructing a new generation of VD profiles for a
representative sample of Galactic GCs.

In Section \ref{obs} we describe the observations and data reduction
procedures.  Section \ref{KA} is devoted to the description of the
kinematic analysis, including a number of tests about the performances
of KMOS (Sect. \ref{test}), the discussion of the determination of the
radial velocities of individual stars (Sect. \ref{RVC}), and the
presentation of the derived VD profile (Sect. \ref{VD}). Discussion
and conclusions are presented in Section \ref{disc}.

\section{Observations and Data Reduction}
\label{obs}

KMOS is a second generation spectrograph equipped with 24 IFUs that
can be allocated within a $7.2\arcmin$ diameter field of view.  Each
IFU covers a projected area on the sky of about
$2.8\arcsec\times2.8\arcsec$, and it is sampled by an array of
14$\times$14 spatial pixels (hereafter spaxels) with an angular size
of $0.2\arcsec$ each.  The 24 IFUs are managed by three identical
spectrographs, each one handling 8 IFUs (1-8, 9-16 and 17-24,
respectively).  At the time of the observations discussed here, IFUs
\#13 and \#16 were not usable.  KMOS is equipped with four gratings
providing a maximum spectral resolution R between $\sim$ 3200 and 4200
over the 0.8-2.5 $\mu$m wavelength range.  We have used the YJ grating
and observed in the 1.00-1.35 $\mu$m spectral range at a resolution
R$\approx$3400, corresponding to a sampling of about 1.75 $\rm
\mathring{A}$ pixel$^{-1}$, i.e. $\sim$ 46 km s$^{-1}$ pixel$^{-1}$ at
1.15 $\mu$m.  This instrumental setup is especially effective in
simultaneously measuring a number of reference telluric lines in the
spectra of giant stars, for an accurate calibration of the radial
velocity, despite the relatively low spectral resolution.  An example
of the observed spectra is shown in Figure~\ref{spec}, with a zoom
around 1.15 $\mu$m to show some isolated telluric lines, and around
1.06 and 1.20 $\mu$m to show some stellar features of interest.

The data presented here have been acquired during the KMOS SV, with
four different pointings on NGC 6388.  The total on-source integration
time for each pointing was 3-5min and it has been obtained with three
sub-exposures of 60-100s each, dithered by $0.2\arcsec$ for optimal
flat-field correction. The typical signal-to-noise ratio (SNR) of the
observed spectra is $\gtrsim$ 50.  We used the ``nod to sky'' KMOS
observing mode and nodded the telescope to an off-set sky field at
$\approx 6\arcmin$ North of the cluster center, for a proper
background subtraction.

The spectroscopic targets have been selected from near-IR data
acquired with SOFI at the ESO-NTT \citep{valenti07}, based on the star
position in the color-magnitude diagrams (CMD) and the radial
distribution within the cluster.

We selected targets with $J < 14$ mag (in order to always have SNR $> 50$)
and sufficiently isolated, without stars brighter than 15 mag within $1\arcsec$ from
their center. We then used ACS-HST data in the $V$ and $I$ bands,
from \citet{lan07}, \citet{sarajedini07} and \citet{dale08}, to
identify additional stars not present in the SOFI catalog.

\begin{figure*}[ht]
\centering
\includegraphics[scale=1.0,angle=270]{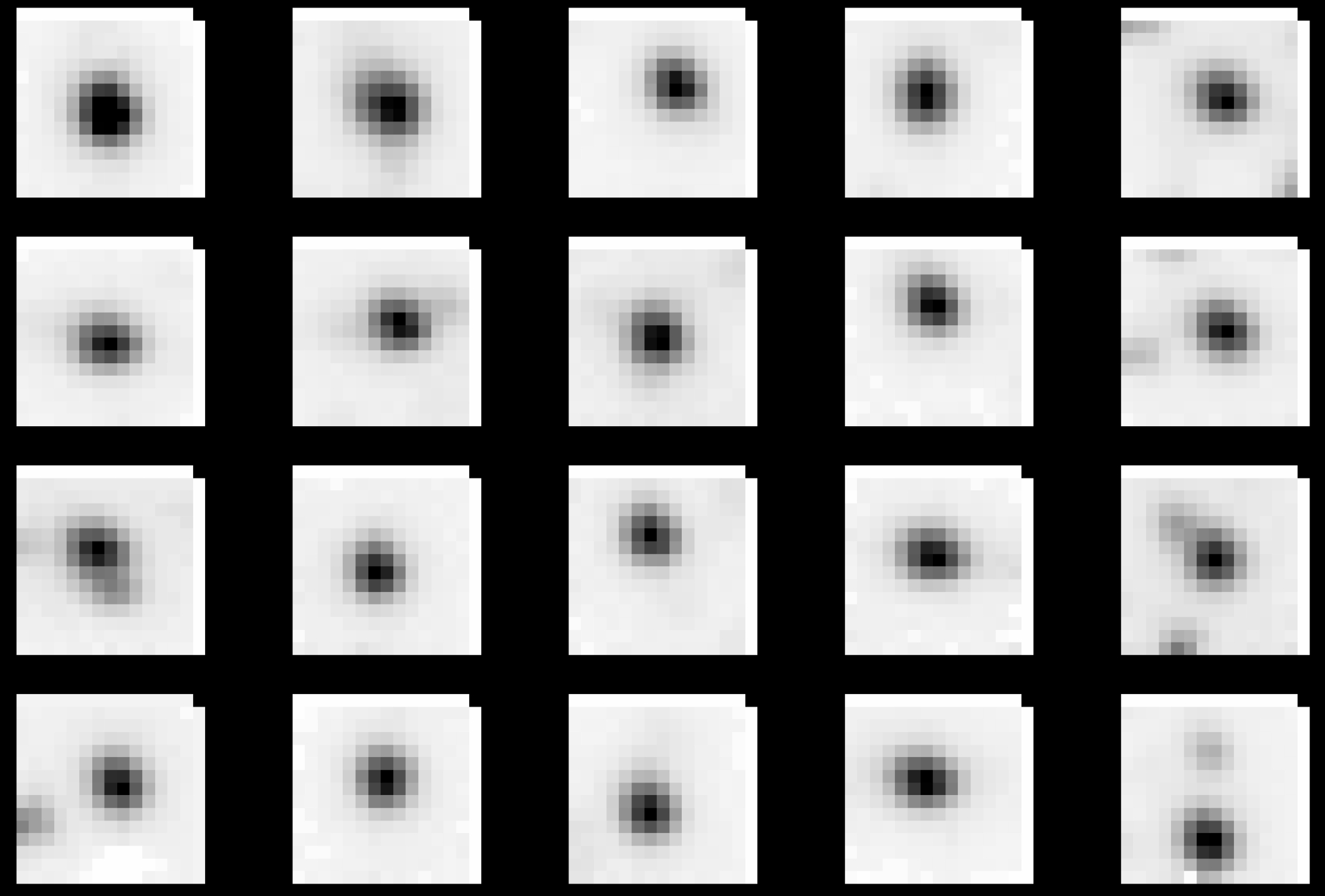}
\caption{Reconstructed images of 20 IFUs obtained during the
first pointing.  In some cases, other stars with a sufficient
spectral SNR and not contaminated by the main target can be
recovered from the same IFU.}
\label{recifu}
\end{figure*}

The raw data have been reduced using the KMOS pipeline version 1.2.6,
which performs background subtraction, flat field correction and
wavelength calibration of the 2D spectra.  The 1D spectra have been
extracted manually by visually inspecting each IFU and selecting the
spectrum from to the brightest spaxel in correspondence of each target
star centroid, in order to minimize the effects of possible residual
contamination by nearby stars and/or by the unresolved stellar
background. An example of the reconstructed images of the
stars observed during the first pointing is shown in Figure \ref{recifu}.

We measured a total of 82 giant stars located within $\sim 70\arcsec$
from the center of NGC 6388. Figure 2 shows the position of the
targets in the the $(I, V-I)$ and $(J, J-K)$ CMDs, while
Figure \ref{map} displays their location in the RA and Dec plane.
Identification number, coordinates and magnitudes
of each target are listed in Table \ref{tabifutot} (the complete
version of the table is available in electronic form). Twelve stars
have been observed twice for cross-checking measurements from
different pointings/exposures.  Seven stars are in common with the
FLAMES-VLT radial velocity sample of L13. In a few cases,
within a single KMOS IFU we could extract the spectra of more
than one star and measure their radial velocity (see Figure \ref{recifu}).

\section{Kinematic Analysis}
\label{KA}

To accurately measure the radial velocities of the observed targets we
made use of cross-correlation techniques with template spectra (in
particular we used the IRAF task FXCOR).  As telluric template we used
a high resolution spectrum of the Earth's telluric
features\footnote{Retrieved from
  \url{http://www.eso.org/sci/facilities/paranal/decommissioned/isaac/tools/spectroscopic$\_$standards.html}.},
convolved at the KMOS YJ grating resolution.  As stellar templates we
used synthetic spectra computed with the TURBOSPECTRUM code
\citep{alvarez98, pletz12}, optimized for cool giants.
We used a set of average templates with photospheric parameters representative
of those of the observed stars and [Fe/H] = --0.5 dex,
the metallicity of NGC~6388 (see \citealt{harris96}, 2010 edition).
For a given star, we also checked the impact of using a different template
with varying the temperature by $\Delta T_{eff}$ $\pm$ 500 K and
the gravity by $\Delta log~g$ $\pm$ 0.5 dex and we verified that it 
has a negligible effect on the final radial velocity measurements ($<$ 1 km s$^{-1}$).

\subsection{Accuracy of the wavelength calibration}
\label{test}

With the purpose of quantifying the ultimate accuracy of the radial
velocity measurements of individual giant stars in crowded fields, we
performed a number of tests aimed at checking the reliability and
repeatability of the wavelength calibration of each KMOS IFU.  Since
KMOS is mounted at a Nasmyth focus and rotates, some flexures are
expected, with impact on the overall spatial and especially spectral
accuracy of the reconstructed 2D spectra.

The KMOS Data Reduction Software (DRS) pipeline allows to take
calibration exposures at several rotator angles and to choose the
frames with the rotator angle closest to the one of the input science
frame and eventually interpolate.  The KMOS DRS pipeline has also the
option of refining the wavelength solution by means of the observed OH
lines. We reduced the spectra by selecting all these options to obtain
the best possible accuracy in the spectral calibration.

However, residual velocity shifts in different spaxels of a given IFU,
as well as in different IFUs, are still possible.  In order to measure
these residual shifts, we selected the spectral region between 1.14
and 1.16 $\mu$m (see Figure~\ref{spec}) containing telluric lines
only, and we cross-correlated the observed spectra from five different
spaxels in a given IFU with the telluric template.  The five spaxels
are the ones where the star centroid is located (having the highest
signal) and the four surrounding (cross-shape) spaxels.  We then
computed the residual wavelength/velocity shifts of the four
surrounding spaxels with respect to the central one used as
reference. As an example, Figure~\ref{ifu15tell} shows the results for
IFU \#15: four different stars (\#245899, \#250977, \#198027 and
\#132259) have been observed in four pointings.  The measured zero
point shifts are normally well within $\pm$10 km s$^{-1}$,
corresponding to 1/4 of a pixel at the spectral resolution of the KMOS
YJ band\footnote{At this resolution, one pixel corresponds to $\sim
  46$ km s$^{-1}$.}, with average values of a few km s$^{-1}$ and
corresponding dispersions within 10 km s$^{-1}$ (see
Table~\ref{tabifu15}).  For each IFU, we finally combined the spectra
from the five spaxels, by using the IRAF task SCOMBINE, and we
measured the radial velocity in the resulting combined one.  The
obtained values (see Figure~\ref{ifu15tell}) are fully consistent with
the average values from individual spaxels (see Table~\ref{tabifu15}).

For the same four stars observed with IFU \# 15, we also used five,
isolated stellar lines in two spectral regions centered at 1.06 and
1.20 $\mu$m (see Figure~\ref{spec}) to compute the velocity shifts.
Also in this case, for each star we extracted the spectrum of the
spaxel with the highest signal and the spectra of the surrounding
(cross-shape) spaxels.  We cross correlated them with suitable
synthetic spectra having photospheric parameters as those of the
target stars.  The resulting velocity shifts with respect to the
central reference spaxel have been plotted in Figure~\ref{ifu15stel},
while the average values are listed in Table~\ref{tabifu15}.  The
inferred average values and dispersions are fully consistent (at
better than 1/10 of a pixel) with those obtained measuring the
telluric lines.  For each star, we finally combined the spectra from
the five spaxels as done in the first test and we measured the radial
velocity in the resulting spectra.  The obtained values (see
Figure~\ref{ifu15tell}) are fully consistent with the average values
from individual spaxels (see Table~\ref{tabifu15}), as well as with
the shifts measured with the telluric lines.

We repeated the same tests by using other stars observed by different
IFUs in different pointings.
As an example, Table \ref{tabifu15} also reports the average shifts for other three stars,
namely $\#$124271, $\#$325164 and $\#$216954 observed by the IFUs $\#$20, $\#$11 and $\#$2 and
collected during the pointings \#1, \#2 and \#4, respectively.
We found values very similar to those derived for the IFU $\#$15,
thus ensuring that the overall wavelength
calibration provided by the KMOS DRS pipeline is normally accurate and
stable in time at a level of a fraction (on average, within 1/10) of a
pixel, both within each IFU and among different IFUs.

The tests performed so far, by using both telluric and stellar lines, 
have demonstrated that the  velocity shifts between the spectra extracted from the spaxel 
with the highest signal and those obtained by combining the spectra from
the cross-shape spaxels are fully consistent one to each other,
thus we decided to use the spectra of the brightest spaxel only,
which is definitely dominated by the target light.
Hence, as a final wavelength calibration check, we selected the
spectrum corresponding to the spaxel with the highest signal in each
observed star. We then cross-correlated this spectrum with the
telluric template as reference, and computed the residual velocity
shift. Figure~\ref{ifutot} shows the results for the active KMOS
IFUs. We find that for a given IFU, the residual velocity shifts as
measured in different stars observed during different
pointings/exposures are normally consistent to each other with an
average dispersion of 3.4 km s$^{-1}$.  Such a dispersion is
relatively small, taking into account that the four exposures on
NGC~6388 were obtained with KMOS at very different rotation angles
with respect to the Nasmyth axis (261$\rm ^o$, 188$\rm ^o$, 326$\rm
^o$, and 97$\rm ^o$ in pointing \#1, \#2, \#3, and \#4, respectively),
indicating that the KMOS optimized calibration procedure is effective
in correcting the effects of spectral flexures.

Since the 24 IFUs of KMOS are managed by three separate spectrographs,
one can also compute the mean shift of each spectrograph, by averaging
the mean shifts from IFUs \#1 to \#8, \#9 to \#16, and \#17 to \#24,
respectively.  We find very similar residual velocity shifts, of
$\approx$-2 km s$^{-1}$ and dispersion of 3-5 km s$^{-1}$.

\subsection{Radial velocity measurements}
\label{RVC}

The tests described in Sect.~\ref{test} indicate that the wavelength
calibration provided by the KMOS pipeline is well suited for
kinematic studies of extragalactic sources.  However, for
precise radial velocity measurements of individual stars, it is
necessary to refine the calibration, by correcting each spectrum for
the corresponding residual velocity shift, as inferred from the
telluric lines. 

Once corrected for such a residual shift, the radial velocity of each
star was computed by cross-correlating the observed spectra with
suitable synthetic ones.  We finally applied the heliocentric
correction by using the IRAF task RVCORRECT.
The final radial velocity errors have been computed from
the dispersion of the velocities derived from each line divided by
the number of lines used (that is $eV_{r}$ = $\sigma / \sqrt{N_{lines}}$).
The average uncertainty in the velocity estimates is 2.9 km s$^{-1}$.

Figure \ref{rvtot} shows the inferred radial velocities as a function
of the radial distance from the cluster center and the histogram of their distribution.
Table \ref{tabifutot} lists the radial velocity values and
corresponding errors, as well as the KMOS IFU and pointing reference
numbers.  For the twelve stars observed twice, the average value of
the measured radial velocities has been adopted. For these stars the measured radial
velocities are in excellent agreement, with an average difference of
$\sim 1$ km s$^{-1}$ between two different exposures and a dispersion
of 3.6 km s$^{-1}$. 

Only seven stars have been found in common with the FLAMES sample
of L13, and the inferred radial velocities from the KMOS spectra turn out to be
in good agreement with the FLAMES ones. In fact by applying a $2-\sigma$ 
rejection criterion, an average difference of
$\langle \Delta \rm V_{KMOS-FLAMES} \rangle$ = --0.2 km s$^{-1}$
($\sigma$ = 2.2 km s$^{-1}$) is found.

We computed the systemic velocity of the KMOS sample by conservatively
using all stars with radial velocities between 60 and 105 km s$^{-1}$,
as done in L13 for the SINFONI and FLAMES samples.
In this velocity range, 75 stars are counted, representing 91$\%$ of the entire KMOS sample.
We found $81.3 \pm 1.5$ km s$^{-1}$, in very good agreement with the value of $82.0 \pm
0.5$ km s$^{-1}$ found by L13 and indicating that all samples are
properly aligned on the same radial velocity scale.

\subsection{Line-of-sight rotation and velocity dispersion profiles}
\label{VD}
To compute the projected rotation and VD profiles from the measured
radial velocities of individual stars we adopted the same approach
described in L13. All the 82 KMOS targets have been considered as
cluster members, since they all have radial velocities between 50 and
130 km s$^{-1}$, which has been adopted as cluster membership
criterion in L13.

To study the possible presence of a rotation signal, we restricted the
analysis to the sample of targets providing a symmetric coverage of
the surveyed area, namely 52 stars located between $9\arcsec$ and
$40\arcsec$ from the centre. For further increasing the sample size,
we took into account 6 additional stars in the same radial range from
the L13 data-set.  We then used the method described in \citet[][and
 references therein; see also L13]{bel12}. No significant rotation
signal has been found from this sample.  Interestingly, however, the
distribution of radial velocities for stars within $20\arcsec$ from
the cluster centre is clearly bimodal (see Figure \ref{rvtot}), thus
suggesting the possible presence of ordered rotation. Unfortunately
only 23 stars have been measured within this radial range and more
data are needed before drawing any firm conclusion about 
ordered rotation in the central regions of NGC 6388 (see also L13).

To compute the VD profile we used the entire KMOS sample (but the two
innermost targets at $r<5\arcsec$ have been conservatively excluded)
and divided the surveyed area in three radial bins,
each containing approximately the same number of
stars: namely $9\arcsec\le r\le 23\arcsec$ (29 stars), $23\arcsec\le
r\le 43\arcsec$ (26 stars), and $43\arcsec\le r\le 70\arcsec$ (25
stars).  The values obtained are $12.9 \pm 2.0$ km s$^{-1}$, at an
average distance of $16\arcsec$, $12.8 \pm 1.9$ km s$^{-1}$ at
$r=33\arcsec$, and $12.2\pm 1.9$ km s$^{-1}$ at $r=56\arcsec$.
The errors have been estimated by following \citealt{pryor93}.
The corresponding profile is plotted in Figure \ref{vdisp}.

\section{Discussion and Conclusions}
\label{disc}

The velocity dispersion values obtained with KMOS are presented in Figure
\ref{vdisp}. We have included for comparison the measurements obtained with SINFONI
and FLAMES from L13 and those derived from integrated-light spectra by \citet{lut11}.
We note that the outermost point of the KMOS VD profile well matches the innermost
FLAMES measure of L13. At the same time, the innermost point of the
KMOS profile is also consistent with the most external value of \citet{lut11}.
Overall, the three new KMOS measurements allow us to sample the velocity
dispersion profile in the spatial region between 9$\arcsec$ and 70$\arcsec$,
and better define the knee of the distribution around 40$\arcsec$ from the
cluster center.

Unfortunately, both crowding and mechanical constraints did not allow
us to allocate more than 1-2 KMOS IFUs per pointing in the very
central region, i.e. at $r\le 9\arcsec$.
Hence, given the limited amount of observing time
during the SV run, only a few stars have been measured in the
innermost region, preventing us to compute a precise VD value closer
to the center.

The final velocity dispersion profile of NGC 6388, obtained from the combination of the
entire sample (namely, SINFONI, KMOS and FLAMES spectra) is presented
in Table \ref{final_vd} and shown in Figure \ref{vdisp_final}.

The results presented here demonstrate the effectiveness of an IFU
facility to perform multi-object spectroscopy of individual stars even
in dense stellar systems and at a modest spectral resolution.  The
KMOS deployable IFU is especially useful in studying the internal
kinematics of GCs for a number of reasons: 1) it allows to sample
stars over a rather large and tunable field of view, according to the
cluster central density and extension; 2) it allows to measure
individual stars and their surroundings, avoiding slit losses and
best-accounting for possible blending effects due to crowding and
unresolved stellar background; 3) it covers a rather wide spectral
range in a single exposure, to simultaneously record stellar features
and telluric lines, and measure accurate radial velocities even at a
spectral resolution R~$\approx$3000.


\acknowledgements The research is part of the project Cosmic-Lab
(http://www.cosmic-lab.eu) funded by the {\it European Research
Council} (under contract ERC-2010-AdG-267675).  E.V. acknowledges
ESO DGDF (SL) 14/51/E. NSO/Kitt Peak FTS data used here were produced
by NSF/NOAO.
We warmly thank the anonymous referee for his/her suggestions in improving the paper.


\begin{deluxetable}{cccccccccrc}
\tablewidth{0pt}
\tablecaption{Individual giant stars in NGC 6388 observed with KMOS.}
\tablewidth{0pt} 
\tablehead{ 
\colhead{star \#} & 
\colhead{RA (2000)}&
\colhead{Dec (2000)}&
\colhead{$V$}&
\colhead{$I$}&
\colhead{$J$}&
\colhead{$K$}&
\colhead{$V_r$}&
\colhead{$eV_r$}&
\colhead{IFU \#}&
\colhead{pointing} 
}
\startdata
 28786 & 264.0777424 & -44.7539232 &  15.12 &  13.32 &  11.91 &  10.84 &  99.7 &  2.1 & 18 &  4 \\ 
 43138 & 264.0633117 & -44.7526742 &  15.85 &  14.19 &  12.90 &  11.93 &  75.2 &  2.4 & 23 &  1 \\ 
 43163 & 264.0631292 & -44.7516067 &  15.64 &  13.86 &  12.48 &  11.42 &  86.8 &  4.8 &  4,19 &  2,3 \\ 
 78741 & 264.0976030 & -44.7395056 &  14.94 &  12.57 &  10.81 &   9.54 &  75.5 &  1.4 &  1,11 &  2,3 \\ 
 81865 & 264.0950571 & -44.7433792 &  15.94 &  14.25 &  12.78 &  11.81 &  83.7 &  6.2 & 10 &  3 \\ 
 93646 & 264.0870297 & -44.7361283 &  15.40 &  13.42 &  11.92 &  10.72 &  94.1 &  2.6 &  6 &  1 \\ 
 94946 & 264.0863051 & -44.7387495 &  15.09 &  13.12 &  11.57 &  10.39 &  83.5 &  2.8 &  5 &  1 \\ 
 96560 & 264.0849040 & -44.7377027 &  14.52 &  12.41 &  10.89 &   9.63 &  79.5 &  1.8 &  2 &  2 \\ 
100296 & 264.0711060 & -44.7360611 &  15.80 &  13.94 &  11.63 &  10.31 &  70.5 &  4.3 &  3 &  3 \\ 
101667 & 264.0820265 & -44.7355902 &  16.16 &  14.67 &   0.00 &   0.00 &  84.4 &  5.2 &  3 &  2 \\ 
.......... \\
\enddata                       
\tablecomments{Identification number, coordinates, optical and NIR
  magnitudes, radial velocities ($V_r$) and errors ($eV_r$) in km
  s$^{-1}$, KMOS IFU and pointing numbers (two values are marked for the twelve
  targets observed twice). The full table is available in the online
  version of the paper.}
\label{tabifutot}
\end{deluxetable}

\begin{deluxetable}{crrcc}
\tablewidth{0pt}
\tablecaption{Results of the wavelength calibration tests.}
\tablewidth{0pt}
\tablehead{ 
\colhead{star \#} & 
\colhead{$\langle \Delta v_{cross}^{telluric}\rangle$} & 
\colhead{$\langle\Delta v_{cross}^{stellar}\rangle$} &
\colhead{IFU \#} &
\colhead{pointing}
}
\startdata
245899 & $+0.7 (8.2)$ & $-1.0 (9.2)$ & 15 & 1\\
250977 & $-0.6 (5.5)$ & $+2.2 (5.7)$ & 15 & 2\\
198027 & $+2.7 (9.5)$ & $-0.1 (11.8)$ & 15 & 3\\
132259 & $+1.0 (4.1)$ & $+0.3 (7.1)$ & 15 & 4\\
 & & & &\\
124271 & $+0.3 (8.2)$ & $+0.7 (9.7)$ & 20 & 1\\ 
325164 & $+0.4 (8.2)$ & $+1.3 (11.5)$ & 11 & 2\\ 
216954 & $-0.1 (5.7)$ & $-0.5 (7.8)$ & 2 & 4\\ 
\enddata                       
\tablecomments{Average velocity shifts and dispersion (in bracket)
among different spaxels distributed in cross-shaped matrixes, with
respect to the reference central spaxel.  Velocity in km s$^{-1}$.}
\label{tabifu15}
\end{deluxetable}

\begin{deluxetable}{rrrcrr}
\tablecaption{Velocity dispersion profile of NGC 6388}
\tablewidth{0pt} 
\tablehead{ 
\colhead{$r_i$} & 
\colhead{$r_e$} &
\colhead{$r_m$} &
\colhead{$N_\star$} &
\colhead{$\sigma_P$} &
\colhead{$e_{\sigma_P}$}}
\startdata
  0.2 &   1.9 &   1.1 & 51 & 13.40 & 1.38 \\ 
  9.0 &  40.0 &  24.1 & 58 & 13.40 & 1.39 \\ 
 40.0 &  75.0 &  55.9 & 57 & 12.80 & 1.23 \\ 
 75.0 & 130.0 & 100.2 & 81 & 11.30 & 0.90 \\ 
130.0 & 210.0 & 164.2 & 84 &  9.10 & 0.70 \\ 
210.0 & 609.0 & 318.8 & 67 &  6.70 & 0.59 \\ 
\enddata                       
\tablecomments{The final profile has been obtained from the combined
  sample of SINFONI, KMOS and FLAMES spectra. The three first columns
give the internal, external and mean radii (in arcseconds) of each
considered radial bin ($r_m$ is computed as the average distance from
the centre of all the stars belonging to the bin), $N_\star$ is the
number of star in the bin, $\sigma_P$ and $e_{\sigma_P}$ are the
velocity dispersion and its rms error (in km s$^{-1}$), respectively.}
\label{final_vd}
\end{deluxetable}


\begin{figure}[ht]
\includegraphics[scale=0.8]{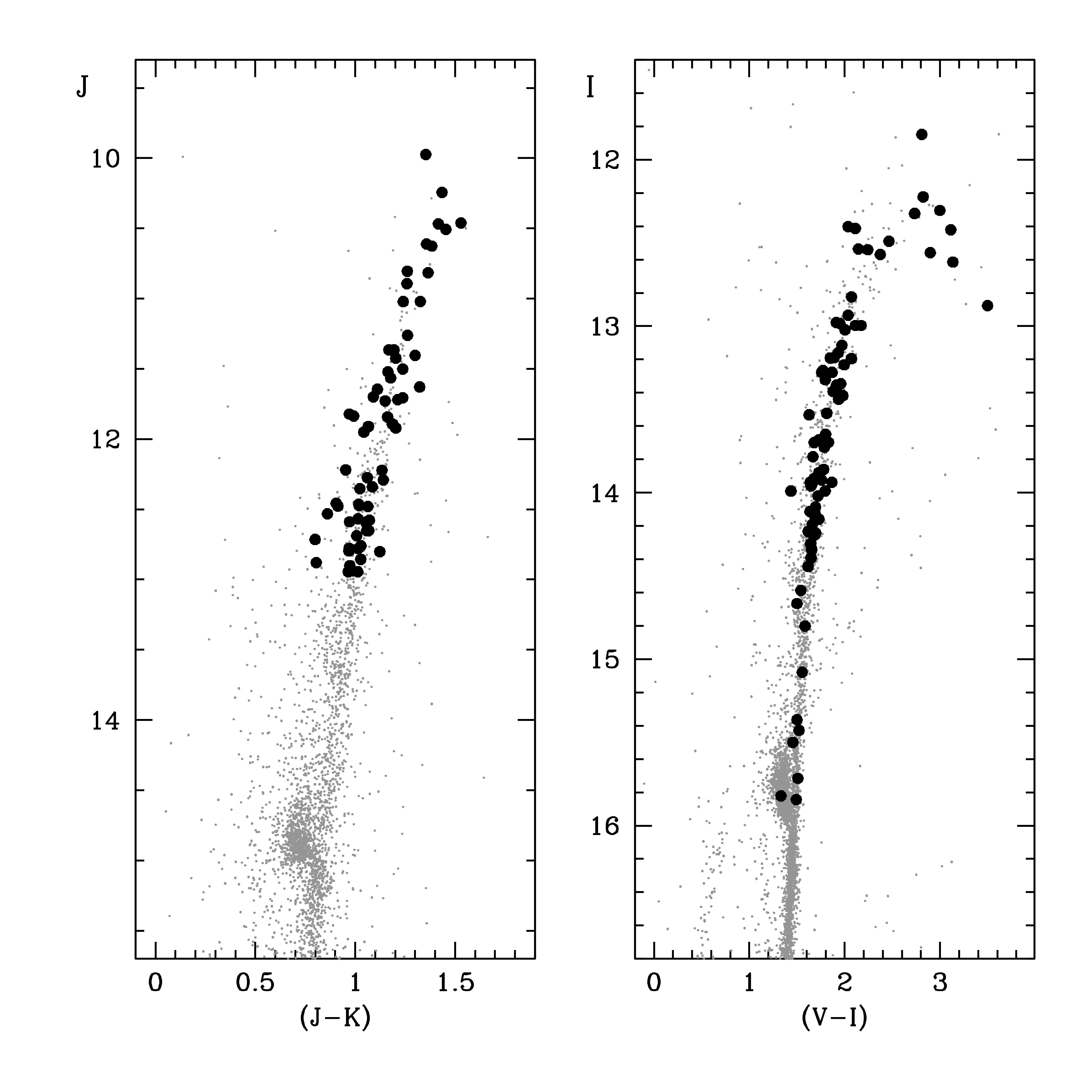}
\caption{The $(J,J-K)$ and $(I,V-I)$ color-magnitude diagrams (left
and right panels, respectively) of NGC 6388, with highlighted the
KMOS targets.}
\label{cmd}
\end{figure}

\begin{figure}[h]
\includegraphics[scale=0.8]{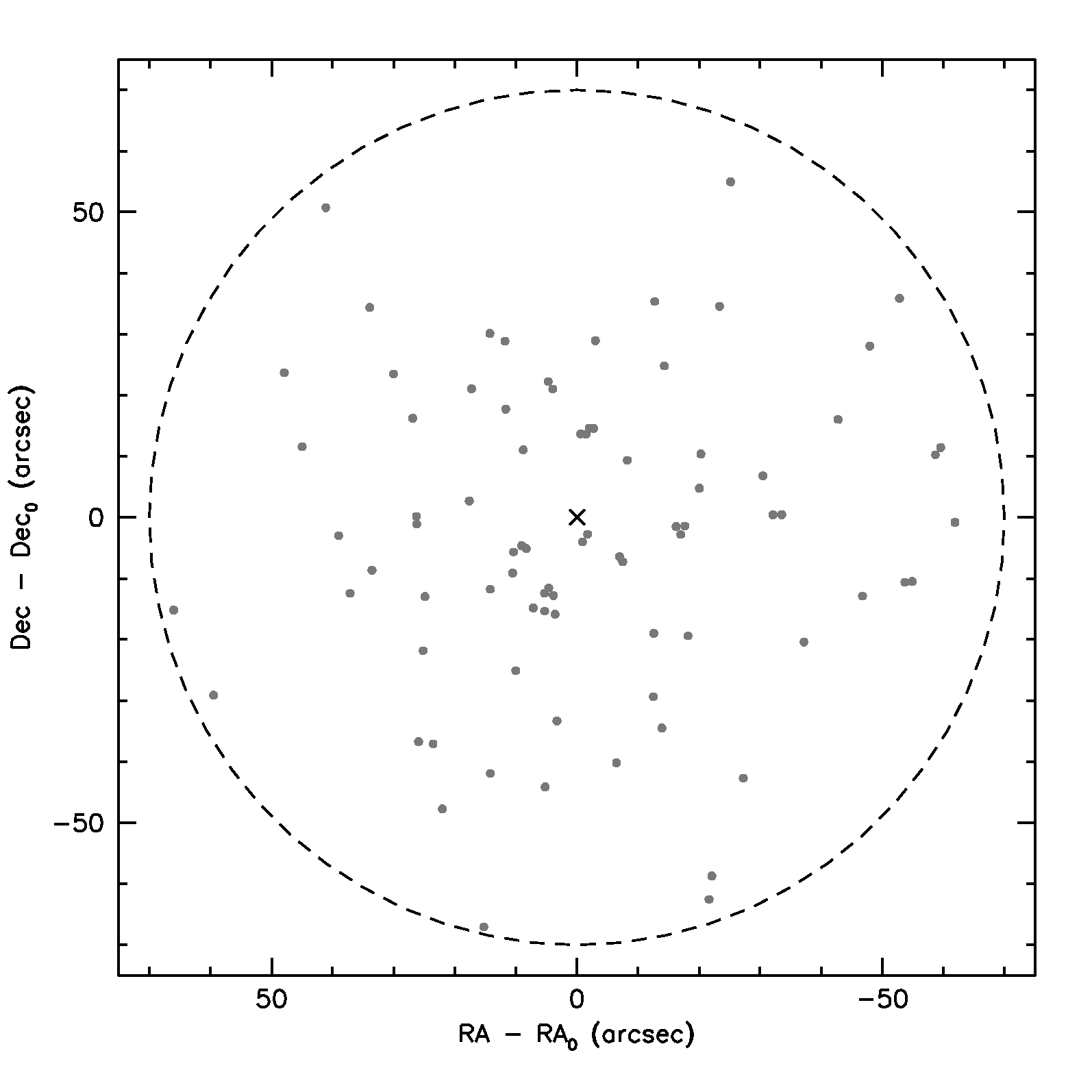}
\caption{The position of the observed target (gray circles) in the central region of NGC~6388.
The black cross marks the center of the cluster as derived in \cite{lan07}
while the black annulus marks a region of $70 \arcsec$ of radius.}
\label{map}
\end{figure}

\begin{figure}[h]
\includegraphics[scale=0.8]{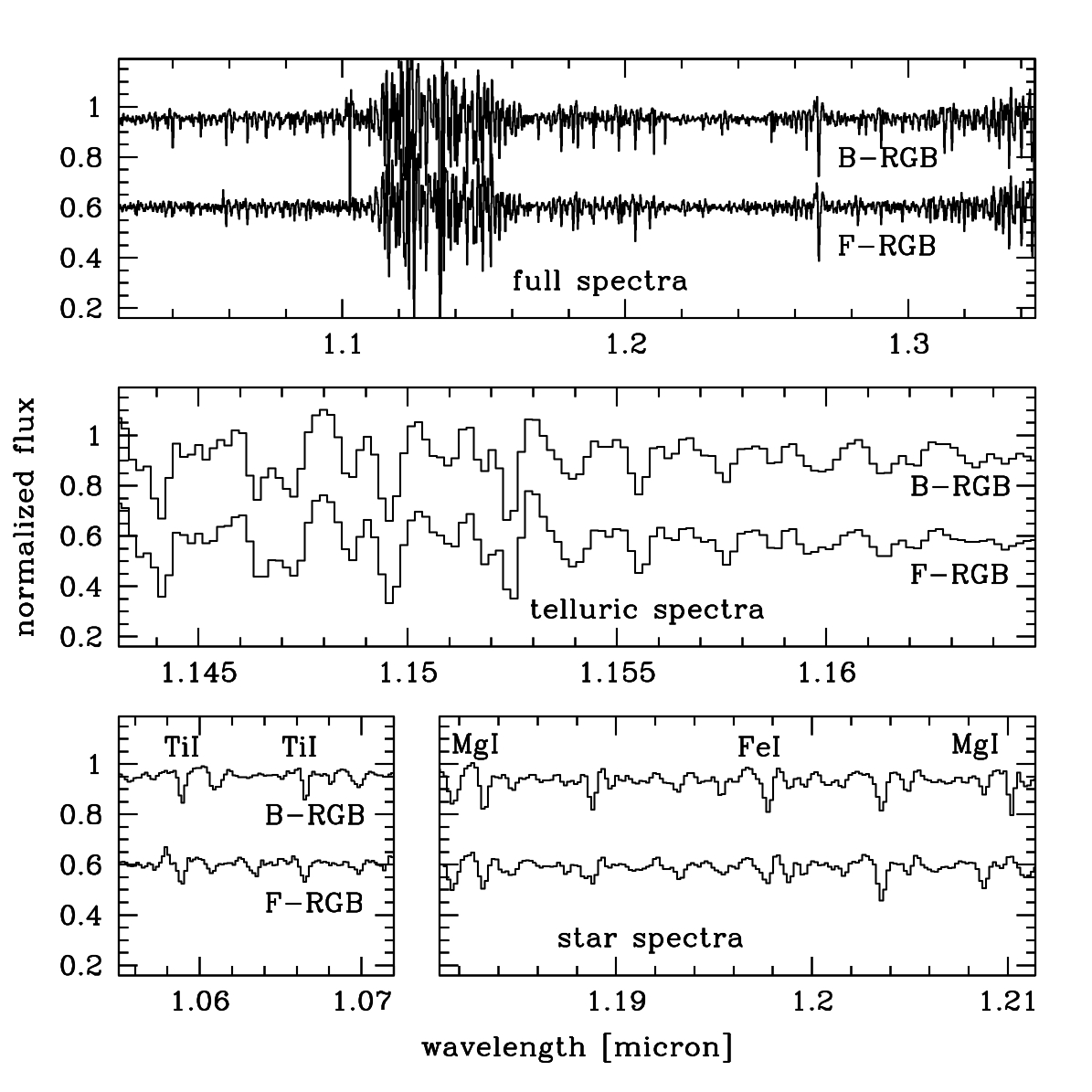}
\caption{An example of the observed KMOS YJ spectra of two giant
stars: a bright B-RGB (star \#245899, J = 10.24 mag, top spectra) and a
faint F-RGB (star \#43138, J = 12.90 mag, bottom spectra). Top
panel: observed spectra. Middle panel: zoomed spectra around 1.15
$\mu$m, including a few isolated telluric lines. Bottom panel:
zoomed spectra around 1.05 and 1.2 $\mu$m, including a few
isolated stellar features of interest.}
\label{spec}
\end{figure}

\begin{figure}[ht]
\includegraphics[scale=0.4]{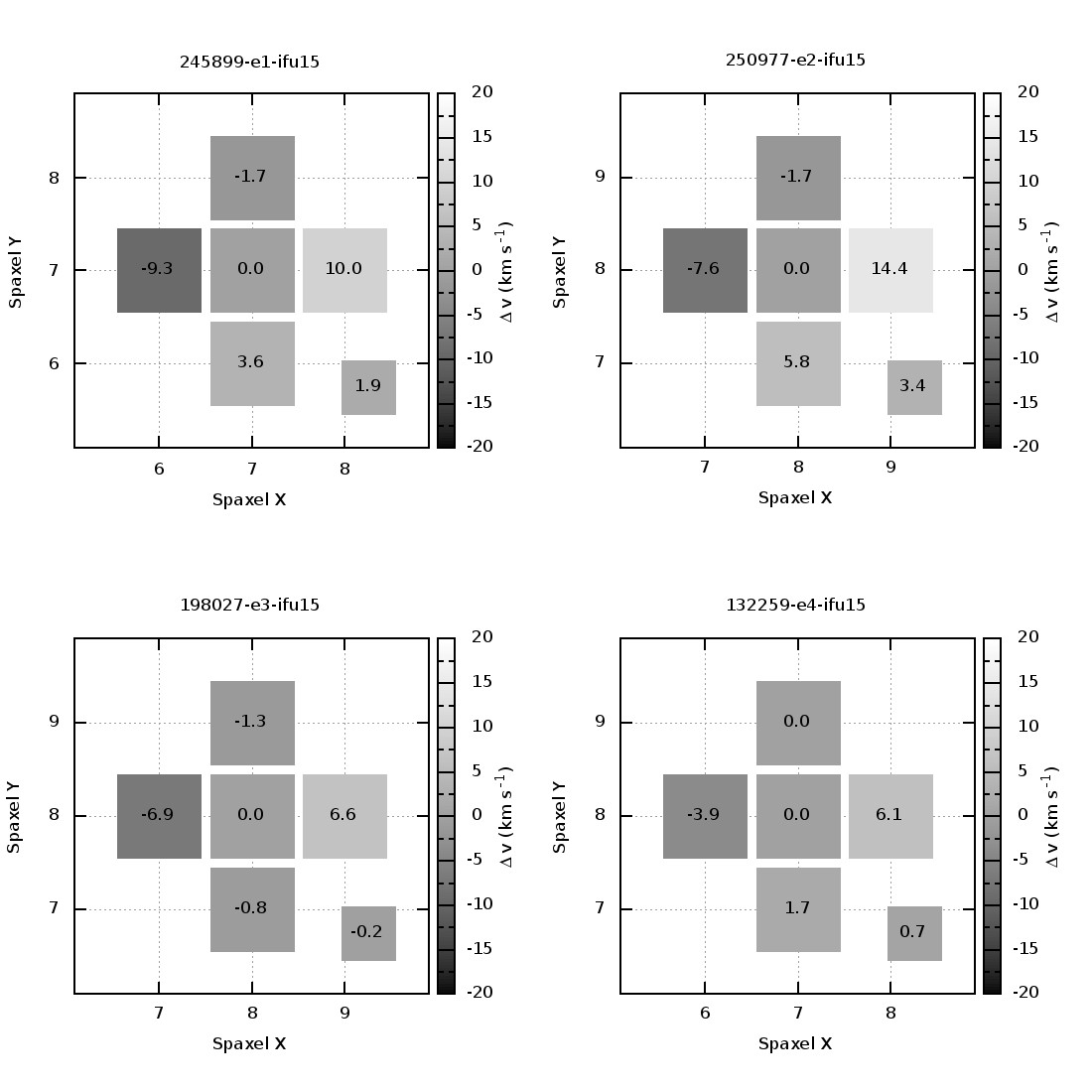}
\caption{The spaxel cross-shaped matrixes used in the wavelength
calibration test. The matrixes are centered on the brightest spaxel
of stars \#245899, \#198027, \#250977 and \#132259 observed with the
IFU \#15 during pointings 1, 2, 3, and 4, respectively.  The number
marked in each spaxel refers to the velocity shift (in km s$^{-1}$)
with respect to the central spaxel, as measured by cross-correlating
telluric lines.  The small square in the bottom-right corner of each
matrix marks the velocity shift with respect to the central spaxel,
as obtained directly from cross-correlating the combined (from the
five spaxels in the cross) spectrum.}
\label{ifu15tell}
\end{figure}

\begin{figure}[ht]
\includegraphics[scale=0.4]{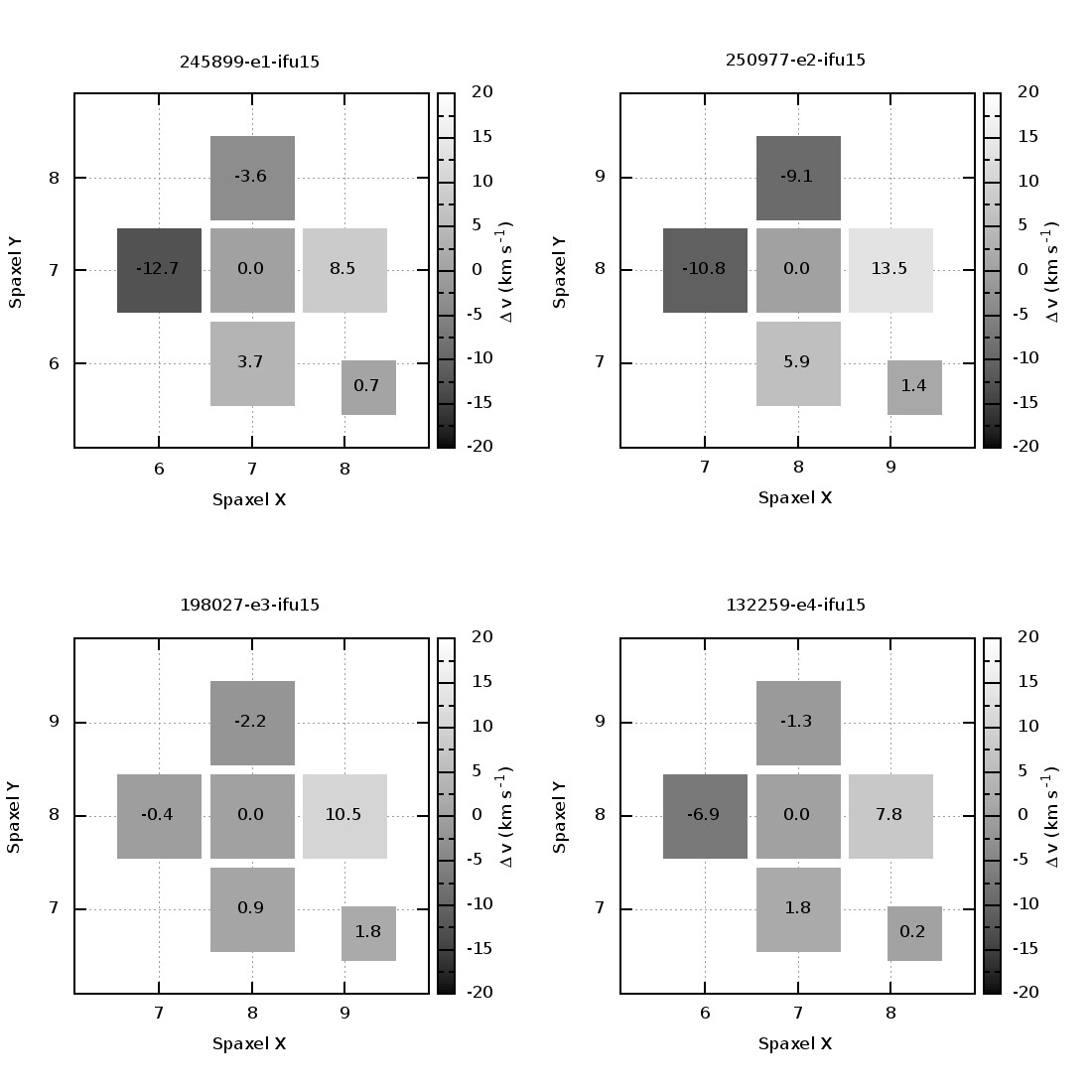}
\caption{As in Figure \ref{ifu15tell}, but for the velocity shifts
measured by cross-correlating stellar lines.}
\label{ifu15stel}         
\end{figure}

\begin{figure}[ht]
\includegraphics[scale=0.8]{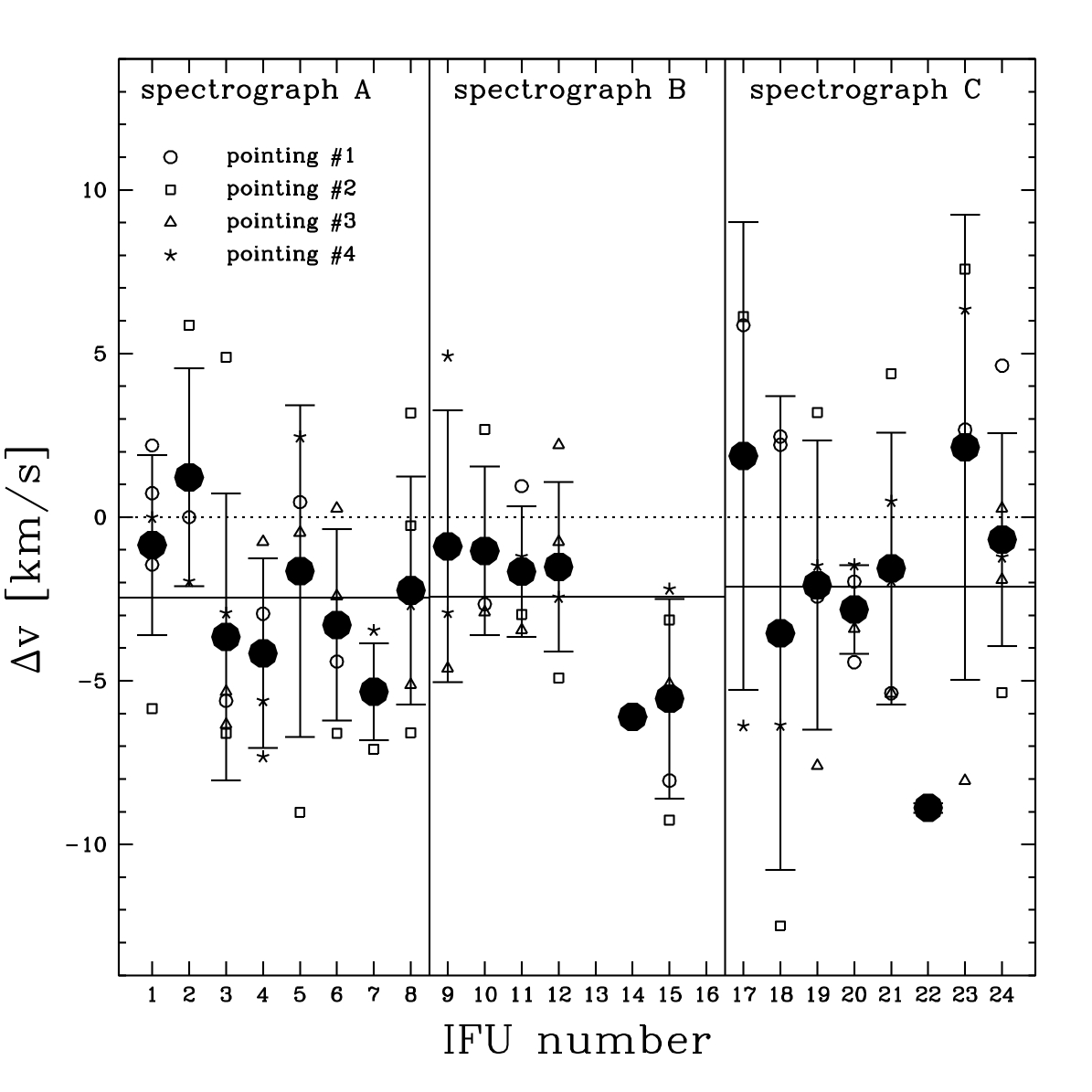}
\caption{The telluric velocity shifts (in units of km s$^{-1}$) of the
spectra extracted from the spaxel with highest signal
for all the target stars observed with the KMOS IFUs (small symbols).
The shifts have been computed by using the telluric template as reference.
The large dots mark the average values and the 1$\sigma$ dispersion as
measured for each IFU, while the horizontal, continuum lines mark
the average values for each of the three spectrographs.
IFUs \#13 and \#16 were not usable during those observations.}
\label{ifutot}
\end{figure}

\begin{figure}[ht]
\includegraphics[scale=0.8]{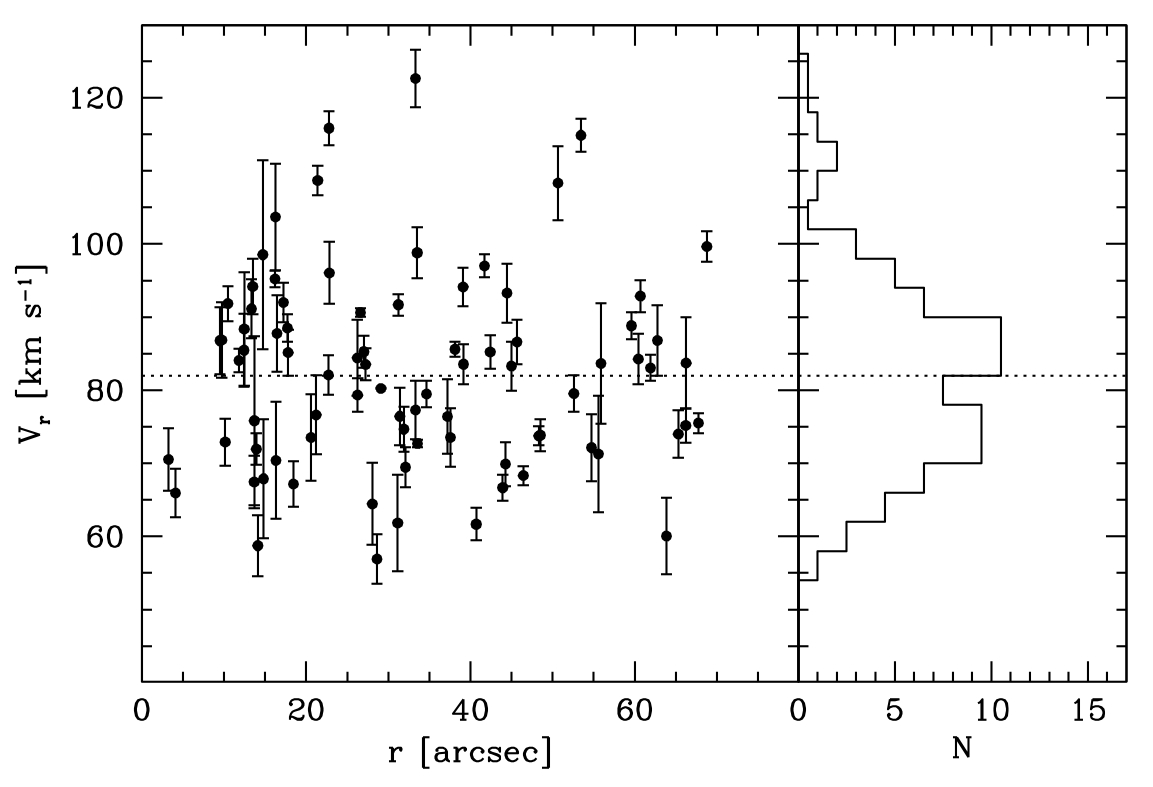}
\caption{Radial velocities as a function of the distance from the
cluster center (left panel) and histogram of their distribution
(right panel), for the 82 stars of NGC 6388 observed with KMOS.
The dotted line marks the systemic velocity of the cluster (82 km s$^{-1}$, from L13).}
\label{rvtot}
\end{figure}

\begin{figure}[ht]
\includegraphics[scale=0.8]{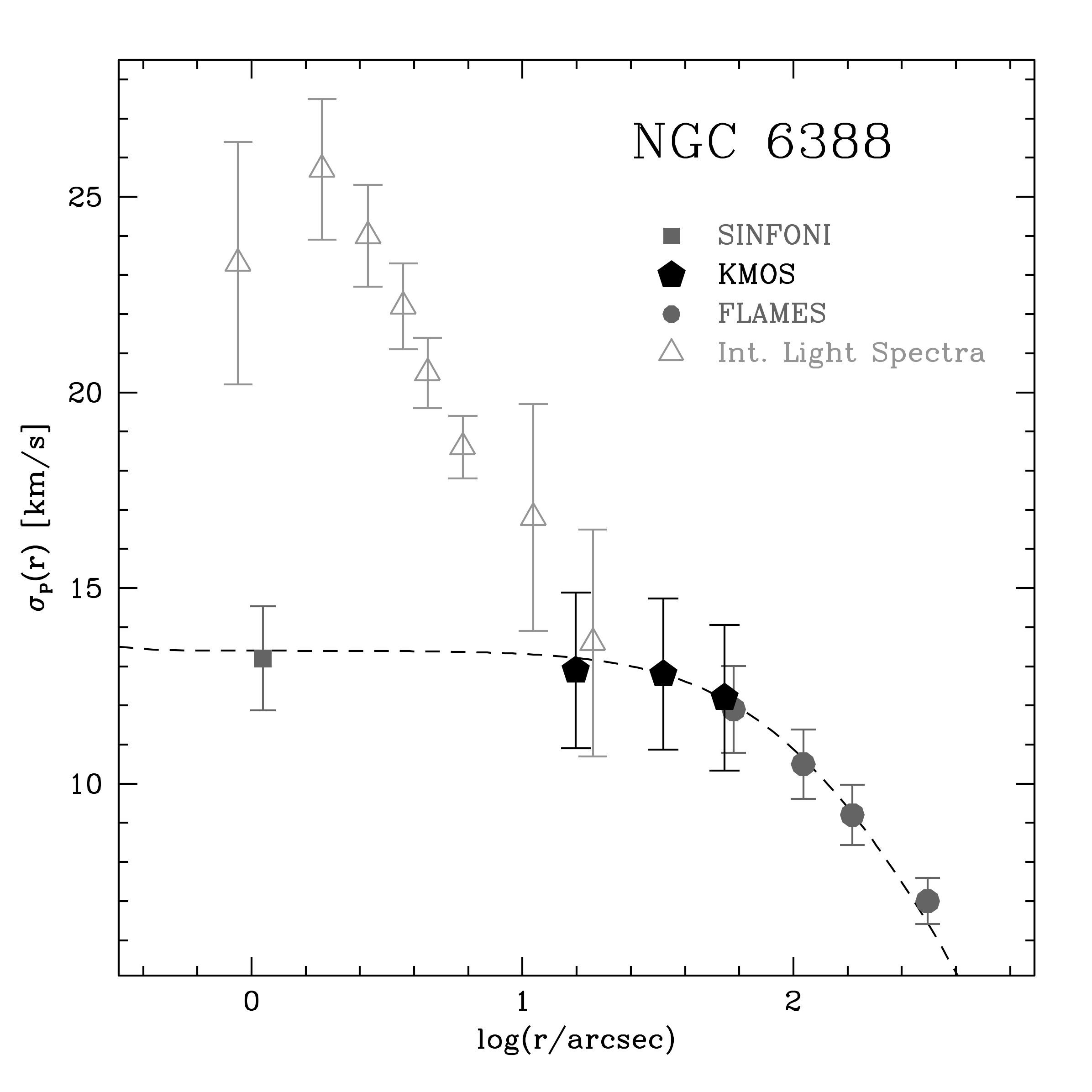}
\caption{Line-of-sight velocity dispersion profile of NGC 6388
computed from the radial velocities of individual stars, as measured
with KMOS (black pentagons, this paper), SINFONI and FLAMES (dark
grey squares and circles, respectively; from L13). The dashed line
correspond to the self-consistent King model plotted in Figure 13 of
L13. The velocity dispersion profile obtained from integrated-light
spectra \citep{lut11} is also shown for comparison (light grey empty
triangles).}
\label{vdisp}
\end{figure}

\begin{figure}[ht]
\includegraphics[scale=0.8]{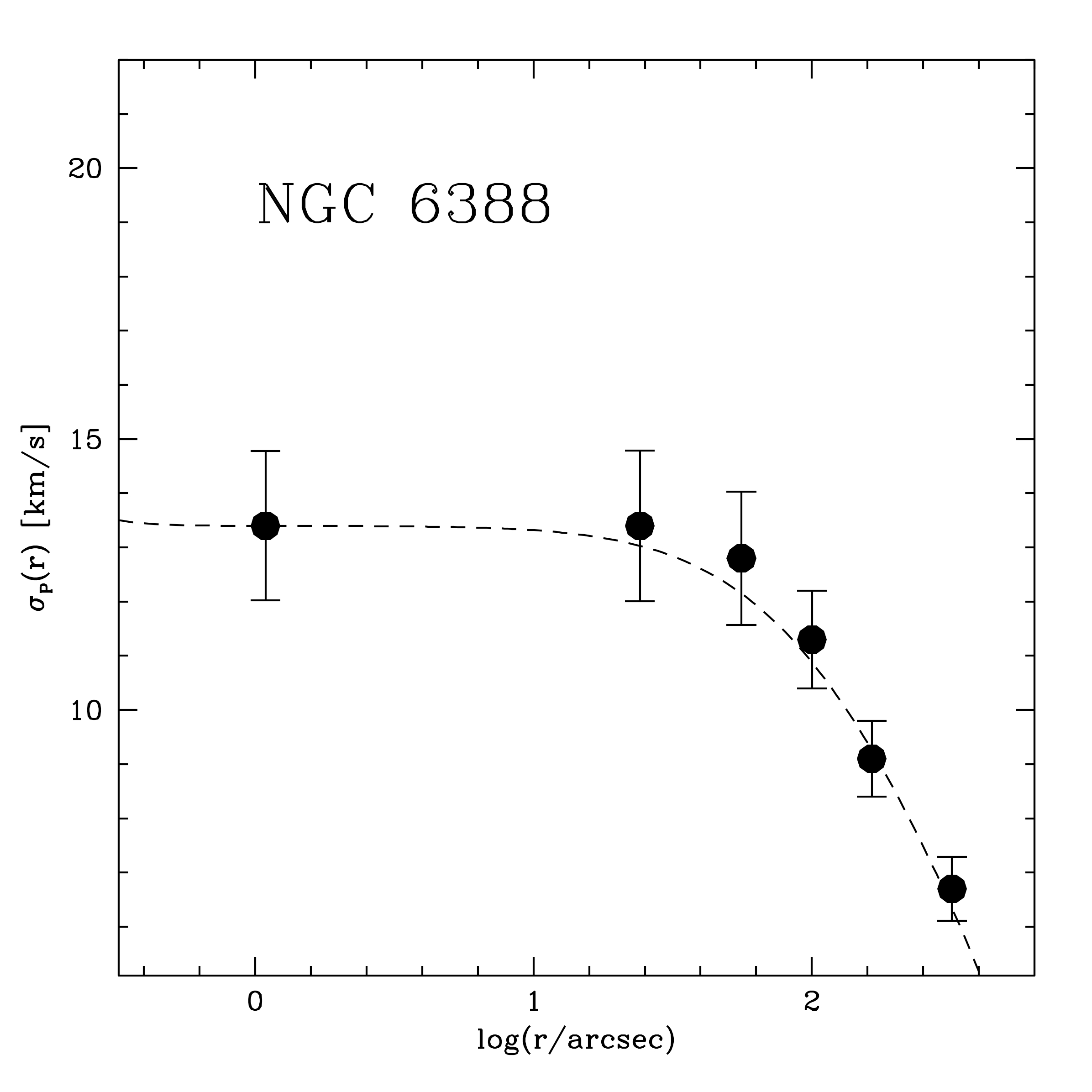}
\caption{Final velocity dispersion profile of NGC 6388, obtained from
the combined sample of individual radial velocities, as measured
from SINFONI, KMOS and FLAMES spectra. The dashed line is as in
Figure \ref{vdisp}.}
\label{vdisp_final}
\end{figure}

\end{document}